\newif\ifen
\newif\ifru
\newcommand{\en}[1]{\ifen#1\fi}
\newcommand{\ru}[1]{\ifru#1\fi}
\entrue

\documentclass[11pt]{article}
\usepackage[utf8x]{inputenc} 
\usepackage{wrapfig} 
\usepackage{graphicx}
\usepackage{hyperref}
\usepackage{appendix}
\usepackage{amsmath}
\usepackage{amssymb}
\usepackage{url}
\usepackage{xcolor}
\ru{
\usepackage[russian]{babel}
}
\en{
\usepackage[english]{babel}
}
\usepackage[left=2cm,right=2cm,top=2cm]{geometry}

\usepackage{natbib}
\setcitestyle{authoryear,open={(},close={)}}

\definecolor{green2}{rgb}{0,0.6,0}

\en{\title{Differential Speckle Polarimetry of Betelgeuse in 2019--2020: \\ the rise is different from the fall}
\author{Safonov B.S.\footnote{safonov@sai.msu.ru}, Dodin A.V., Burlak M.A., Goliguzova M.V., Fedoteva A.A.,\\ Zheltoukhov S.G., Lamzin S.A., Strakhov I.A., Voziakova O.V.}
\date{11th May 2020}
}

\ru{
\title{Дифференциальная спекл-поляриметрия Бетельгейзе: \\ поведение оболочки в минимуме 2019-2020 гг}
\author{Сафонов Б.С.\footnote{safonov@sai.msu.ru}, Додин А.В., Бурлак М.А., Возякова О.В., Голигузова М.В., \\ Желтоухов С.Г., Ламзин С.А., Страхов И.А., Федотьева А.А.}
\date{6 мая 2020}
}

\begin{document}

\maketitle

\begin{abstract}
\ru{
Опубликованные недавно спектры \citep{Levesque2020} и изображения с высоким угловым разрешением \citep{Montarges2020} показывают что глубокий минимум блеска Бетельгейзе 2019-2020 года вызван увеличением содержания пыли в атмосфере. Подробный мониторинг таких событий может быть весьма полезным для построения согласованных моделей механизма потери массы у звезд на поздних стадиях эволюции. В таких наблюдениях принципиально важным представляется применение методов, позволяющих прослеживать неоднородность атмосферы звезды. 

Мы выполнили наблюдения Бетельгейзе методом дифференциальной спекл-поляметрии на 2.5-м телескопе Кавказской Горной Обсерватории ГАИШ МГУ в 17 эпох   охватывающих период минимума на длинах волн 465, 550, 625, 880~нм. Все наблюдения показывают наличие вокруг звезды поляризованной отражательной туманности с угловым размером $\approx0.1^{\prime\prime}$. Морфология туманности изменяется по мере входа в минимум и выхода из него. Полная поляризованная яркость оболочки оставалась постоянной до середины февраля 2020 года, в то время как поток от звезды упал в 2.5 раза. С середины февраля и до начала апреля 2020 года поляризованный поток от оболочки возрос в 2.1 раза, в то время как яркость звезды вернулась к уровню октября 2019 года. Основываясь на этих данных и спектра видимого диапазона, полученного нами 6 апреля 2020 года, при мы подтверждаем вывод о том что минимум вызван образованием пылевого облака локализованного на луче зрения. Количественная характеризация этого облака станет возможна с привлечением данных о его собственном тепловом излучении в ИК--диапазоне.}
\en{Recently published spectral \citep{Levesque2020} and high angular resolution \citep{Montarges2020} observations of Betelgeuse suggest that the deep minimum of 2019-2020 was caused by an enhanced dust abundance in the stellar atmosphere. Detailed monitoring of such events may prove useful for constructing consistent physical models of mass loss by evolved stars. For such observations it is fundamentally important to employ methods resolving an inhomogeneous stellar atmosphere.

We present the differential speckle polarimetric observations of Betelgeuse at 2.5-m telescope of Caucasian Mountain Observatory of SAI MSU covering the period of 2019-2020 minimum. The observations were secured on 17 dates at wavelengths 465, 550, 625 and 880~nm. The circumstellar reflection nebula with the angular size of $\approx0.1^{\prime\prime}$ was detected for all the dates and at all wavelengths. The morphology of the nebula changed significantly over the observational period. Net polarized brightness of the envelope remained constant until February 2020, while the stellar $V$ band flux decreased 2.5 times. Starting from mid-February 2020, polarized flux of the envelope rose 2.1 times, at the same time the star returned to the pre-minimum state of October 2019. Basing on these data and our low resolution spectrum obtained on 2020-04-06 we confirm a conclusion that the minimum is caused by the formation of a dust cloud located on the line of sight. A quantitative characterisation of this cloud will be possible when the data on its thermal radiation are employed.
}

\end{abstract}

\ru{
\section{Введение}}
\en{
\section{Introduction}}

\ru{
С ноября 2019 по март 2020 года Бетельгейзе испытала ослабление блеска с $m_V=0.6$ до $1.6$. Это ослабление хоть и укладывается в картину нерегулярной переменности звезды, но тем не менее было самым глубоким за весь период фотоэлектрических наблюдений \citep{Guinan2020}. }
\en{
During the period from November 2019 to March 2020  Betelgeuse passed through an unusually deep minimum. Its brightness decreased from $m_V=0.6$  to $m_V=1.6$. Although this minimum was the deepest for the whole history of photoelectric observations, it fits the irregular variability of the star \citep{Guinan2020}. }

\ru{
Спектральные наблюдения \citet{Levesque2020} показали, что температура фотосферы изменилась в минимуме незначительно. Авторы этой работы приходят к выводу, что ослабление вызвано возрастанием оптической толщи пыли на луче зрения. Вместе с тем они нашли, что покраснение звезды не изменилось по сравнению с дозатменным состоянием и соответствует $A_{\rm V}=0.62,$ при стандартном законе поглощения, из чего был сделан вывод, что минимум вызван крупными пылинками. Отметим, что возможно и другое объяснение наблюдаемого серого поглощения. В случае неравномерного затмения звездного диска пылевым облаком кажущийся закон поглощения выглядит всегда более серым, чем закон, отвечающий свойствам пылинок \citep{Natta1984}. Признаки неравномерного по звездному диску затмения непосредственно следуют из наблюдений с высоким угловым разрешением на VLT/SPHERE \cite{Montarges2020} 27 декабря 2019 года в полосе 644~нм показали гораздо более асимметричную фотосферу чем годом ранее: южная половина выглядела ослабленной. }
\en{Spectral observations by \citet{Levesque2020} showed that the temperature of stellar photosphere changed during the minimum insignificantly. Authors of that work concluded that the fainting of the star was caused by the increase of dust absorption on the line of sight. At the same time they found out that the reddening of the star has not changed in comparison to pre--eclipse star and corresponded to $A_{\rm V}=0.62,$ for the standard extinction law. This made it possible to make a conclusion that the minimum is caused by large dust particles. We note that an alternative explanation for the grey extinction is possible. In case of inhomogeneous obscuration of the stellar disc by a dust cloud, the apparent extinction law is always more grey, than the intrinsic one \citep{Natta1984}. High angular resolution observations by \citet{Montarges2020} showed much more asymmetric photosphere on 2019-12-27 than a year before: the southern hemisphere looked faint. This observation favours the inhomogeneous obscuration.} 

\ru{
\citet{Cotton2020} привели результаты прецезионной поляриметрии полного потока от объекта. Полная поляризация несколько снизилась в затмении. Авторы списывают это на уменьшившийся вклад поляризованного излучения фотосферы вследствие возрастания оптической толщи на луче зрения. Однако полная поляризация подобных объектов уже чрезвычайно ослаблена усреднением по лимбу. Для полностью симметричной звезды она ожидается равной нулю \citep{Clarke2010}. Таким образом, снижение поляризации в затмении, обнаруженное \citet{Cotton2020} находится в противоречии с наблюдениями \citet{Montarges2020}, показывающими менее симметричную картину фотосферы в этот же период. Все это подчеркивает необходимость проведения не отдельных, а регулярных мониторинговых наблюдений объекта с высоким угловым разрешением, подобных \citep{Montarges2020,Haubois2019}.
}
\en{
\citet{Cotton2020} presented precision aperture polarimetry of the object before the minimum and at the bottom of it. The net polarization decreased somewhat in the eclipse. \citet{Cotton2020} attribute this to decreased input of polarized radiation from the  photosphere due to obscuring material on the line of sight. However, the net polarization of such objects is strongly suppressed by azimuthal averaging. It is expected to be zero for perfectly symmetric star  \citep{Clarke2010}. In this respect fall of polarization found by \citet{Cotton2020} seems to contradict the images by \citet{Montarges2020} showing less symmetrical photosphere during the minimum. This paradox emphasizes the necessity of frequent high--angular resolution polarimetric observations similar to those by \citet{Montarges2020,Haubois2019}.
}

\ru{
Здесь мы представляем наблюдения звезды методом поляризационной интерферометрии на длинах волн 465, 550, 625 и 880~нм, выполненные в 17 дат с конца октября 2019 года по конец апреля 2020 года. Наши наблюдения показывают наличие компактной околозвездной оболочки и позволяют проследить изменения ее яркости и морфологии на протяжении минимума. При интерпретации мы также используем собственный спектр низкого разрешения, снятый в видимом диапазоне на выходе из минимума.}
\en{Here we present the observations of Betelgeuse by means of polarimetric interferometry at wavelengths 465, 550, 625 and 880~nm, conducted on 17 dates from the end of October 2019 until the end of April 2020. We reliably detect a compact circumstellar envelope around the star and follow the changes of its net brightness and morphology during the minimum. For the interpretation we also use our own low--resolution visible--range spectrum obtained  after the minimum.}

\ru{\section{Наблюдения}}

\en{\section{Observations}}

\ru{Наблюдения были выполнены на спекл-поляриметре 2.5-м телескопа КГО ГАИШ МГУ \citep{Safonov2017,Safonov2019a} в 14 эпох в период с 27 октября 2019 года по 24 апреля 2020 года. Также дополнительно мы будем рассматривать наблюдения 2 декабря 2018 года и 23 января 2019 года.}

\en{The observations were secured with the SPeckle Polarimeter (SPP) mounted at the 2.5-m telescope of Caucasian Mountain Observatory of SAI MSU \citep{Safonov2017,Safonov2019a}. These observations cover a period from 2019-10-27 to 2020-04-24. Additionally, we will consider observations made on 2018-12-02 and 2019-01-23.}

\begin{figure}[t!]
\centering
\begin{tabular}{cc}
\includegraphics[width=98mm]{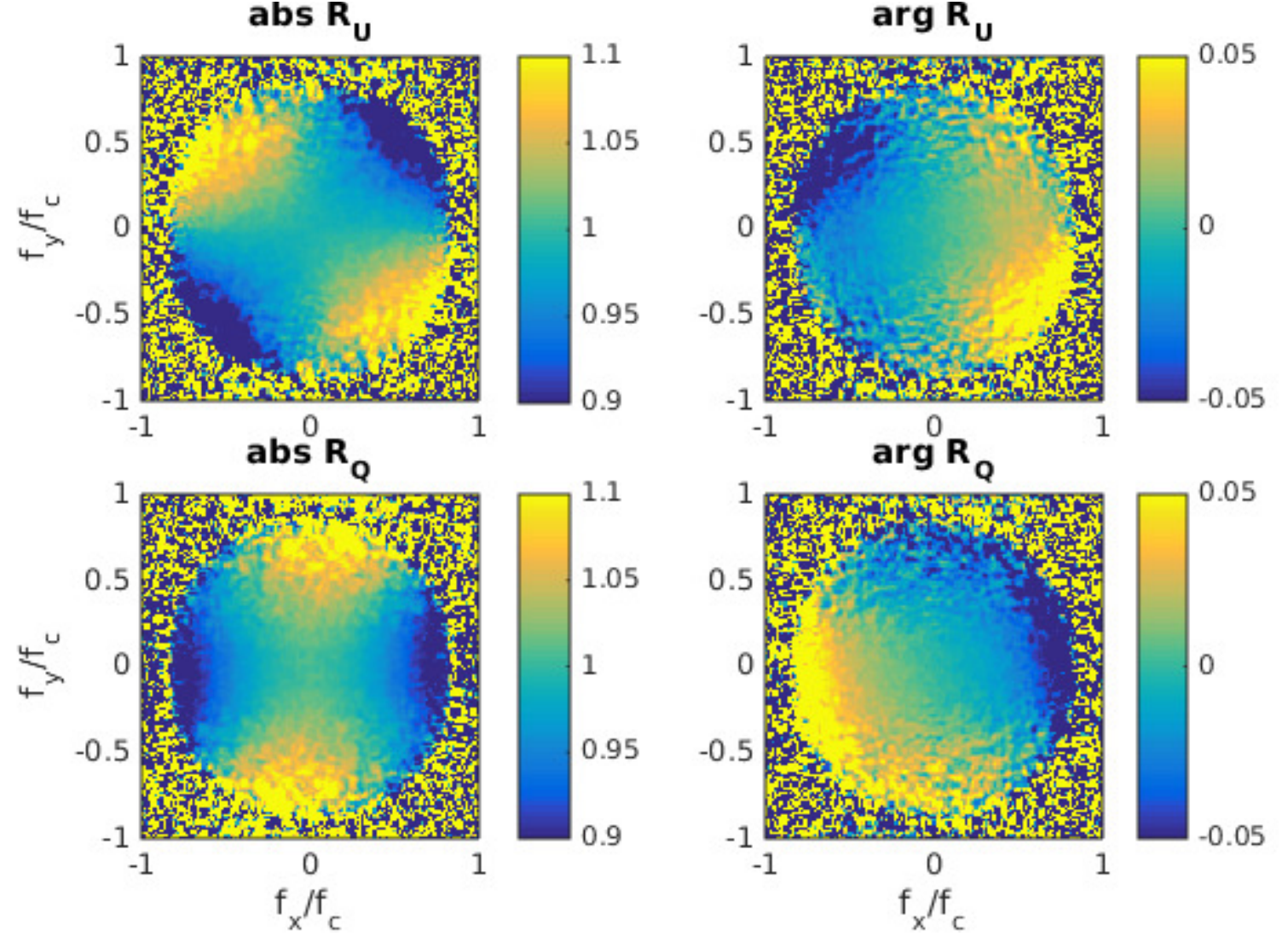} &
\includegraphics[width=78mm]{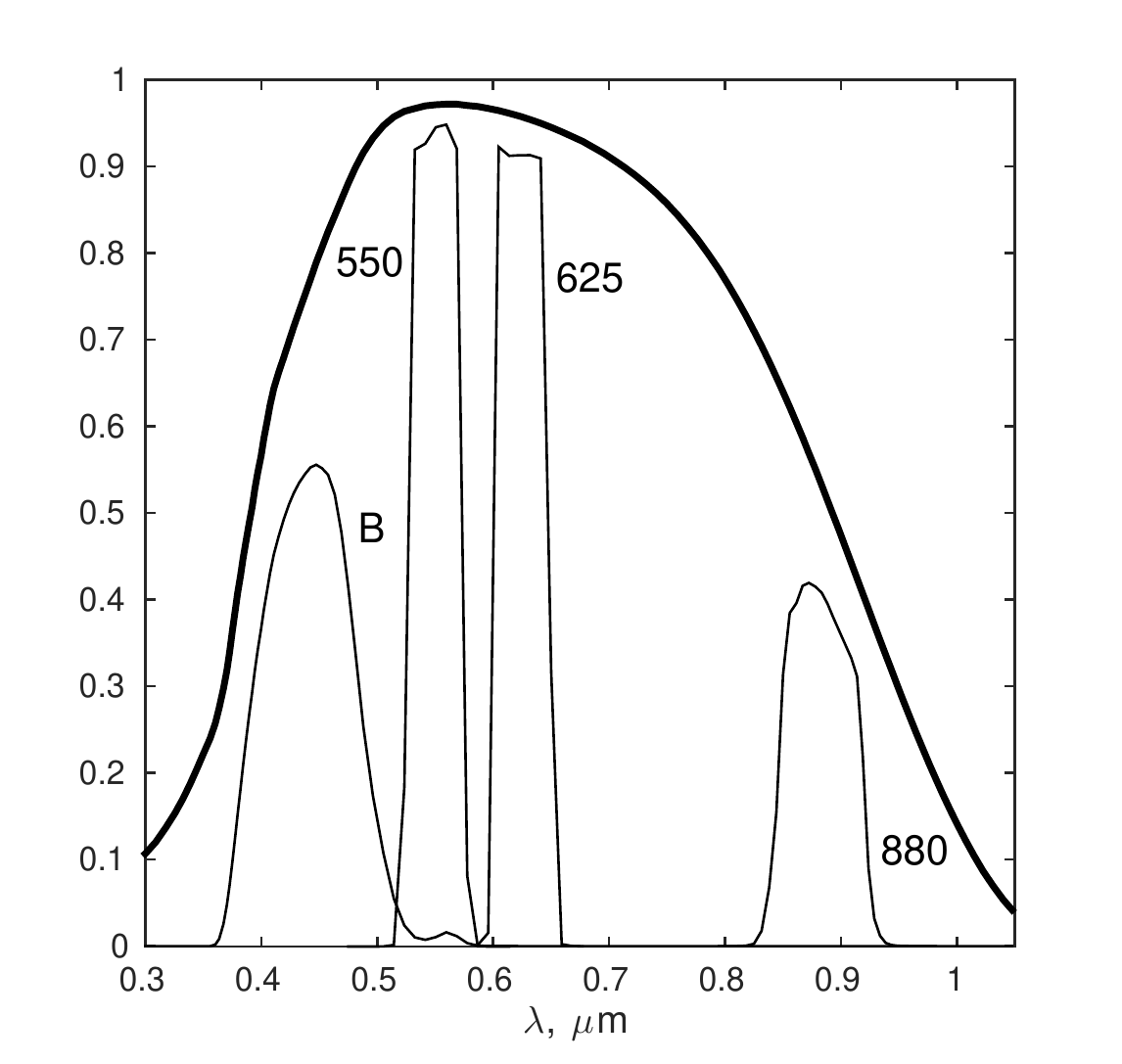} \\
\end{tabular}
\caption{\ru{Слева: пример измерения величины $\mathcal{R}$ для Бетельгейзе. Измерение получено 11 декабря 2019 года в фильтре 550~нм. По осям пространственная частота, нормированная на частоту среза. Север вверху, восток справа.} \en{ Left panel: the measurement of $\mathcal{R}$ for Betelgeuse on 2019-12-11 in band 550~nm. Spatial frequency normalized by cut--off frequency $f_c=D/\lambda$ of the system is along the axes. $D$ --- aperture diameter. North is up, East is left.} 
\ru{Справа: тонкие линии --- фотометрические полосы, примененные в работе, умноженные на квантовую эффективность детектора (толстая линия).}\en{ Right panel: thin lines are the passbands of the filters employed in this work multiplied by the quantum efficiency of the detector Andor iXon 897 (thick line).} 
\label{fig:rval_bands}}
\end{figure}

\ru{СПП представляет собой  спекл-интерферометр, регистрирующий одновременно два ортогонально поляризованных seeing-limited изображения объекта. Прибор также оснащен полуволновой пластинкой, играющей роль модулятора. В процессе накопления серии короткоэкспозиционных изображений HWP вращается непрерывно со скоростью 300$^{\circ}/$сек, обеспечивая swapping of состояний поляризации в фокальной плоскости каждые 150~мс.  Экспозиция составляет 30~мс, типичная длина серии 3000-6000 кадров.  Спектральный диапазон прибора --- от 400~нм до 1100~нм. На рис. \ref{fig:rval_bands}, справа приведены фотометрические полосы в которых были выполнены наблюдения для данной работы. Подробно обстоятельства наблюдений представлены в Табл.~\ref{table:obslog}.}

\en{SPP is a fast high angular resolution camera which acquires two orthogonally polarized seeing--limited images of an object simultaneously. The instrument is equipped with a half--wave plate (HWP) acting as a modulator. During the acquisition of a series of short--exposure images the HWP rotates continuously at an angular speed of 300$^{\circ}$/s, providing the swapping of polarization states in focal plane once each 150 ms. The exposure is 30~ms, the typical length of series is 3000-6000 frames. The wavelength range of the instrument is from 400 to 1100~nm. The passbands of the filters employed in this work are given in Fig.~\ref{fig:rval_bands}. The detailed observational log is presented in  Appendix \ref{app:obslog}.}

\ru{Полученные серии были обработаны методом дифференциальной спекл-поляриметрии (ДСП). Основным результатом такой обработки является отношение видностей объекта в ортогональных поляризациях:}
\en{The obtained series were processed by Differential Speckle Polarimetry (DSP). The basic result of it is the ratio of visibilities in orthogonal polarizations:}
\begin{equation}
\mathcal{R}_Q = \frac{\widetilde{I}-\widetilde{Q}}{\widetilde{I}+\widetilde{Q}},\,\,\,\mathcal{R}_U = \frac{\widetilde{I}-\widetilde{U}}{\widetilde{I}+\widetilde{U}}.
\label{eq:defR}
\end{equation}
\ru{Здесь $\widetilde{I},\widetilde{Q},\widetilde{U}$ --- преобразования Фурье от распределений параметров Стокса. $\boldsymbol{f}$ --- вектор двумерной пространственной частоты. Величина $\mathcal{R}$, которую также иногда называют differential polarimetric visibility, была использована для исследования пылевых оболочек звезд на поздних стадиях эволюции в работах \citet{Norris2012,Safonov2019b,Fedotyeva2020}. В частности, к Бетельгейзе метод был применен  \citet{Haubois2019}. Подчеркнем, что мы рассматриваем как амплитуду, так и фазу $\mathcal{R}$, что значительно повышает чувствительность метода к {\it асимметрии} распределения поляризованного потока. Методика ДСП не требует квазиодновременных наблюдений калибровочных звезд. Большая часть наблюдений была выполнена в фокусе Нэсмита, коррекция за эффекты инструментальной поляризации выполнялась методом, описанным в \cite{Safonov2019a}.}
\en{Here $\widetilde{I},\widetilde{Q},\widetilde{U}$ are the Fourier transforms of Stokes parameters distributions. $\boldsymbol{f}$ is a vector of two--dimensional spatial frequency. The $\mathcal{R}$ value, which is sometimes called differential polarimetric visibility, was used for the studies of dusty envelopes of evolved stars by \citet{Norris2012,Safonov2019b,Fedotyeva2020}. Specifically, \citet{Haubois2019} applied the method to Betelgeuse. We emphasize that we consider both amplitude and phase of $\mathcal{R}$, which greatly increases the sensitivity of the method to {\it asymmetry} of polarized flux distribution. Our method of DSP does not require quasi--simultaneous observations of calibration sources. The majority of observations were conducted while the instrument was mounted at the Nasmyth-2 focal station of the 2.5-m telescope. The correction for instrumental polarization effects was performed as described by \citet{Safonov2019b}.}

\ru{Для звезд, не обладающих поляризованными оболочками $\widetilde{Q}(\boldsymbol{f})=0$ и $\widetilde{U}(\boldsymbol{f})=0$ $\forall\boldsymbol{f}$, следовательно можно ожидать что $\mathcal{R}=1$. Пример измерения $\mathcal{R}$ для Бетельгейзе приведен на Рис.~\ref{fig:rval_bands}. Как видно, фиксируется отклонение модуля $\mathcal{R}$ от единицы, что говорит о наличии поляризованной туманности. Вид $\mathcal{R}$ --- butterfly pattern --- типичен для отражательной туманности \cite{Norris2012,Haubois2019}. Из Рис.~\ref{fig:rval_bands} также видно, что присутствует отклонение аргумента от нуля. Это говорит о значительной асимметрии поляризованного потока. }
\en{For the stars without polarized circumstellar structures $\widetilde{Q}(\boldsymbol{f})=0$ and $\widetilde{U}(\boldsymbol{f})=0$ $\forall\boldsymbol{f}$, therefore $\mathcal{R}=1$ is expected. The example of $\mathcal{R}$ measurement for Betelgeuse is given in Fig.~\ref{fig:rval_bands}. One can tell that  $\mathcal{R}$ significantly deviates from 1, which gives evidence for the existence of a polarized nebula. The butterfly pattern in $\mathcal{R}$ is very typical for a reflection nebula \citep{Norris2012,Haubois2019}. The phase of $\mathcal{R}$ significantly deviates from zero, hinting the asymmetry of polarized flux distribution.}

\ru{В \cite{Safonov2019a} мы описали как величина $\mathcal{R}$, при условии знания ее амплитуды и фазы, может быть использована для восстановления изображения оболочки в поляризованной интенсивности. Соответствующее вычисление выполняется в предположении того что в объекте доминирует точечная и неполяризованная звезда. Бетельгейзе имеет угловой размер, которым уже нельзя пренебречь, более того, согласно \cite{Montarges2020}, звезда значительно несимметрична. В приложении~\ref{app:finitesize} мы показали что это сказывается на изображении, восстанавливаемом методом \cite{Safonov2019a}, незначительно.}
\en{\cite{Safonov2019a} describe how the $\mathcal{R}$ value can be used for the reconstruction of polarized intensity distribution of the envelope, given that both its amplitude and phase is known. This reconstruction implies that the object is dominated by a point--like and unpolarized star. At the same time, Betelgeuse has an angular size which we cannot neglect. Moreover, according to \cite{Montarges2020}, the star lacks the point symmetry. In Appendix~\ref{app:finitesize} we demonstrate that both factors do not affect the reconstructed images much.}

\begin{figure}[p]
\centering
\begin{tabular}{c}
\includegraphics[width=175mm]{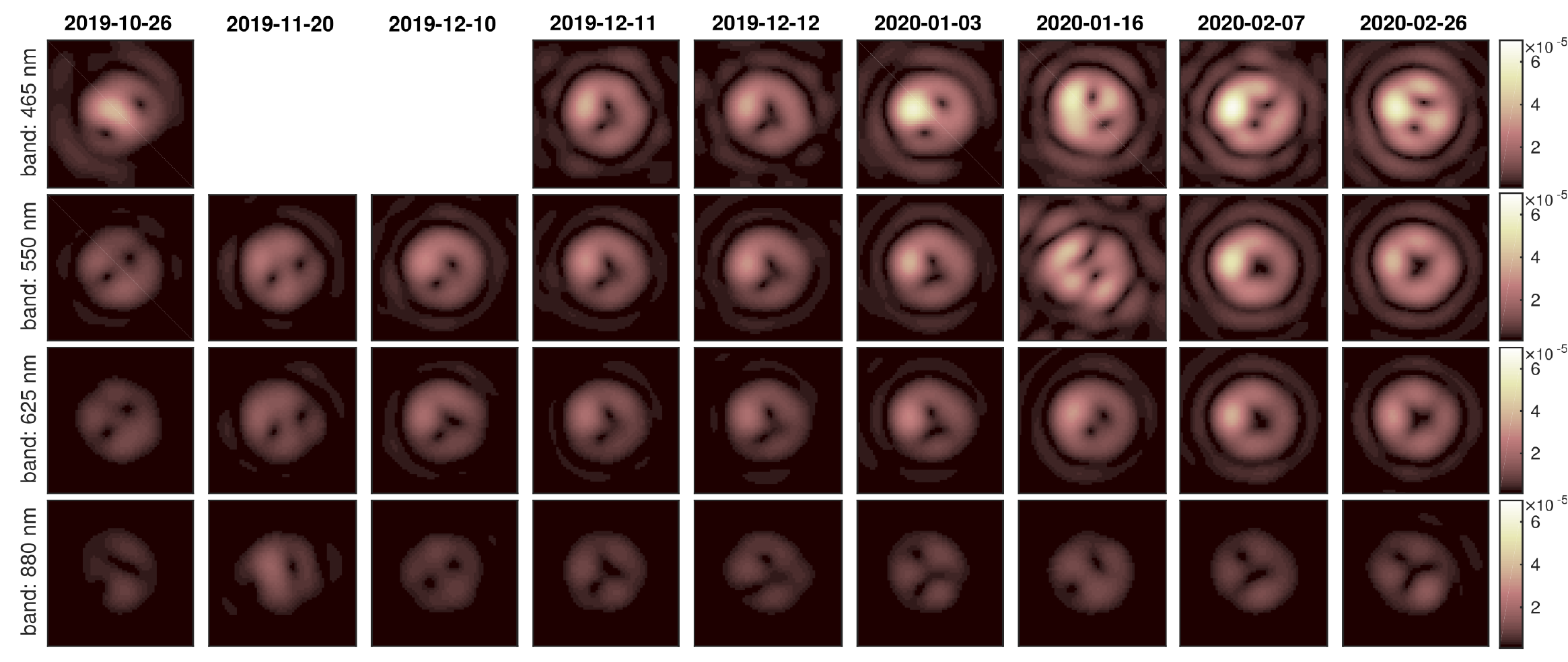} \\
\includegraphics[width=175mm]{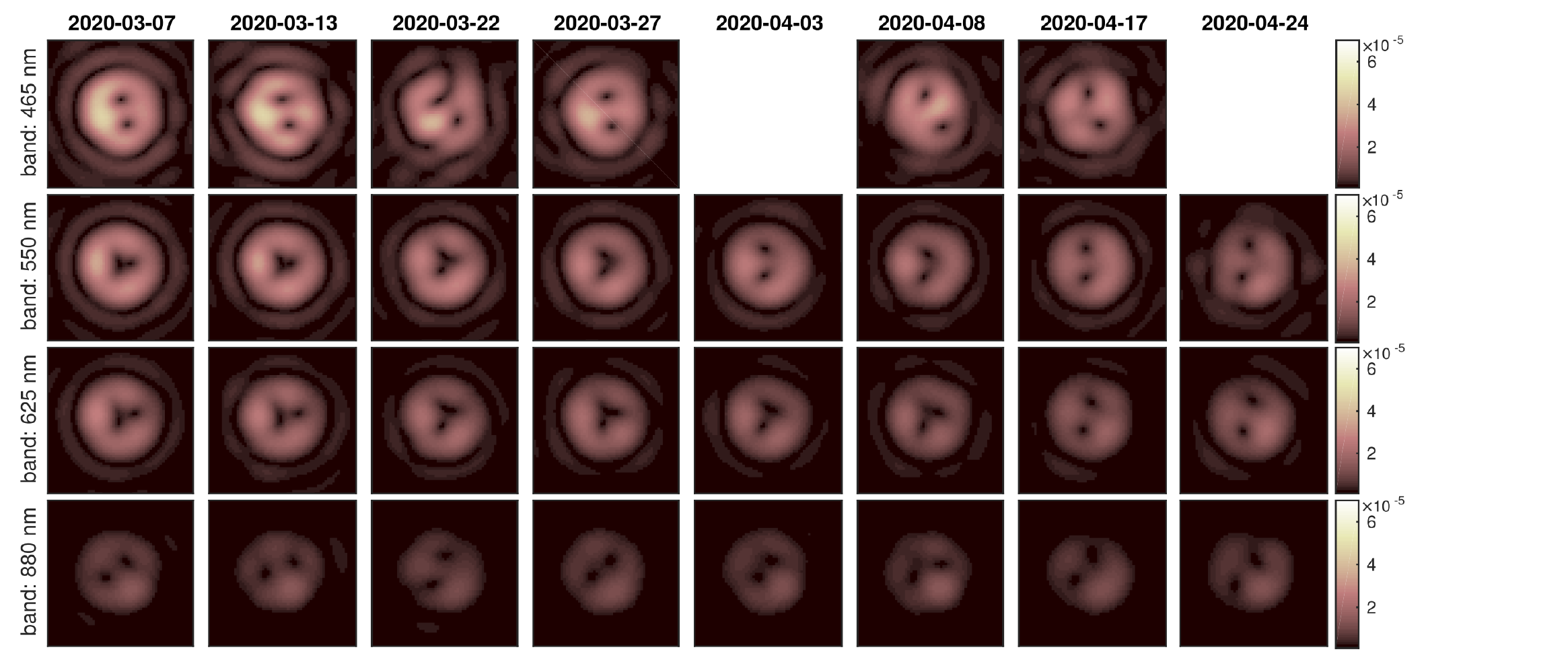} \\
\end{tabular}
\caption{\ru{Изображения околозвездной оболочки Бетельгейзе в поляризованном свете, восстановленные методом, описанным в \cite{Safonov2019a}, некоторые комментарии см. также в приложении~\ref{app:finitesize} данного документа. Размер пикселя 5 mas, что в 4 раза меньше, чем исходный угловой масштаб прибора. Яркость пропорциональна поляризованному потоку приходящемуся на пиксель деленному на полный неполяризованный поток от звезды. Шкала яркостей приведена справа. Размер отображаемой области $0.3\times0.3^{\prime\prime}$, север вверху, восток слева. Строки соответствуют фильтрам, стоблцы --- датам. В некоторые даты наблюдения в фильтре $B$ не выполнялись.}
\en{The images of circumstellar envelope of Betelgeuse in polarized light, reconstructed using the method described by \citet{Safonov2019a}. Some comments are also given in Appendix~\ref{app:finitesize} of this paper. The pixel size is 5~mas, which is 4 times smaller than the initial angular scale of the camera. The brightness is proportional to polarized flux in the pixel relative to net unpolarized flux of the object. The size of displayed region is 0.3$\times$0.3$^{\prime\prime}$, North is up, East is left. Rows correspond to bands, columns --- to dates. On some dates observations in the $B$ band were not conducted.}
\label{fig:images}}
\end{figure}

\ru{На Рис.~\ref{fig:images} приведены изображения, построенные этим методом, для всех полученных наблюдений. Изображения во всех фильтрах были получены сверткой с дифракционной функцией рассеяния точки, соответствующей длине волны $\lambda=880$~нм, что обеспечивает возможность прямого сравнения пространственных масштабов.}
\en{The reconstructed images are presented in Fig.~\ref{fig:images} for all the obtained observations. The images for all the filters were convolved with diffraction--limited point spread function, corresponding to $\lambda=880$, thus allowing for the direct comparison of spatial scales.}

\ru{Во всех восстановленных изображениях наблюдается азимутальная картина поляризации (см. Рис.~\ref{fig:finitesize}) --- плоскость поляризации в каждой точке перпендикулярна направлению на звезду. Поэтому чтобы не перегружать изображения плоскость поляризации на них не приведена. Изображения для трех последовательных дат 10, 11 и 12 декабря дают представление о воспроизводимости результатов. Измерения в полосе 550~нм 15 января 2020 года имеют плохое соотношение сигнал-шум из-за значительных вариаций атмосферной прозрачности.}
\en{The azimuthal pattern of polarization is present in all the images --- the polarization plane is oriented normally to local direction to the star center. Thus, for clarity, the polarization plane is not plotted. The images for three consecutive dates 2019-12-10, 11, 12 provide a measure of the repeatability of the results. The measurement in the 550 filter made on 2019-01-15 has poor signal--to--noise ratio because atmospheric transparency variations were too large.}

\ru{Вспомогательные спектральные наблюдения в видимом диапазоне были выполнены на Transient Double--beam Spectrograph\footnote{\url{http://lnfm1.sai.msu.ru/kgo/instruments/tds}} КГО ГАИШ МГУ 6 апреля 2020 года. Поскольку Бетельгейзе является слишком яркой звездой для этого инструмента, её изображение на щели было расфокусировано, что позволило получать спектры с экспозицией 3-5 сек. Всего было получено 4 снимка. Обработка производилась с помощью специально созданного пакета программ на языке python и включала в себя следующие этапы: вычитание темновых кадров, коррекция кривизны изображения щели и калибровка по длинам волн по спектру Ne-Kr-Pb лампы,  исправление неравномерности чувствительности вдоль щели по источнику с непрерывным спектром, экстракция спектра в апертуре фиксированной ширины, коррекция за кривую реакции системы.}

\en{Auxiliary spectral observations in visible range were conducted with the Transient Double--beam Spectrograph\footnote{\url{http://lnfm1.sai.msu.ru/kgo/instruments/tds}} at the same telescope on 2020-04-06. Due to the extreme brightness of Betelgeuse, we had to defocus its image on the slit in order to obtain 3--5~sec unsaturated exposures. We obtained 4 frames containing spectra. The processing was performed using a specialized pipeline written in python. It consists of the following steps: dark frames subtraction, slit image curvature correction, wavelength calibration using the spectrum of Ne-Kr-Pb light source, spectrum extraction in aperture of fixed width and correction for the response curve.}

\ru{Для целей сравнения со спектром из работы \cite{Levesque2020} усредненный по 4 измерениям спектр был нормирован на континуум и сглажен скользящим средним для уменьшения спектрального разрешения.}
\en{For the comparison with the spectrum by \citet{Levesque2020} we averaged our data over 4 measurements, normalized it by continuum and smoothed by running average to match the spectral resolution.}

\begin{figure}[b!]
\begin{center}
\includegraphics[width=140mm]{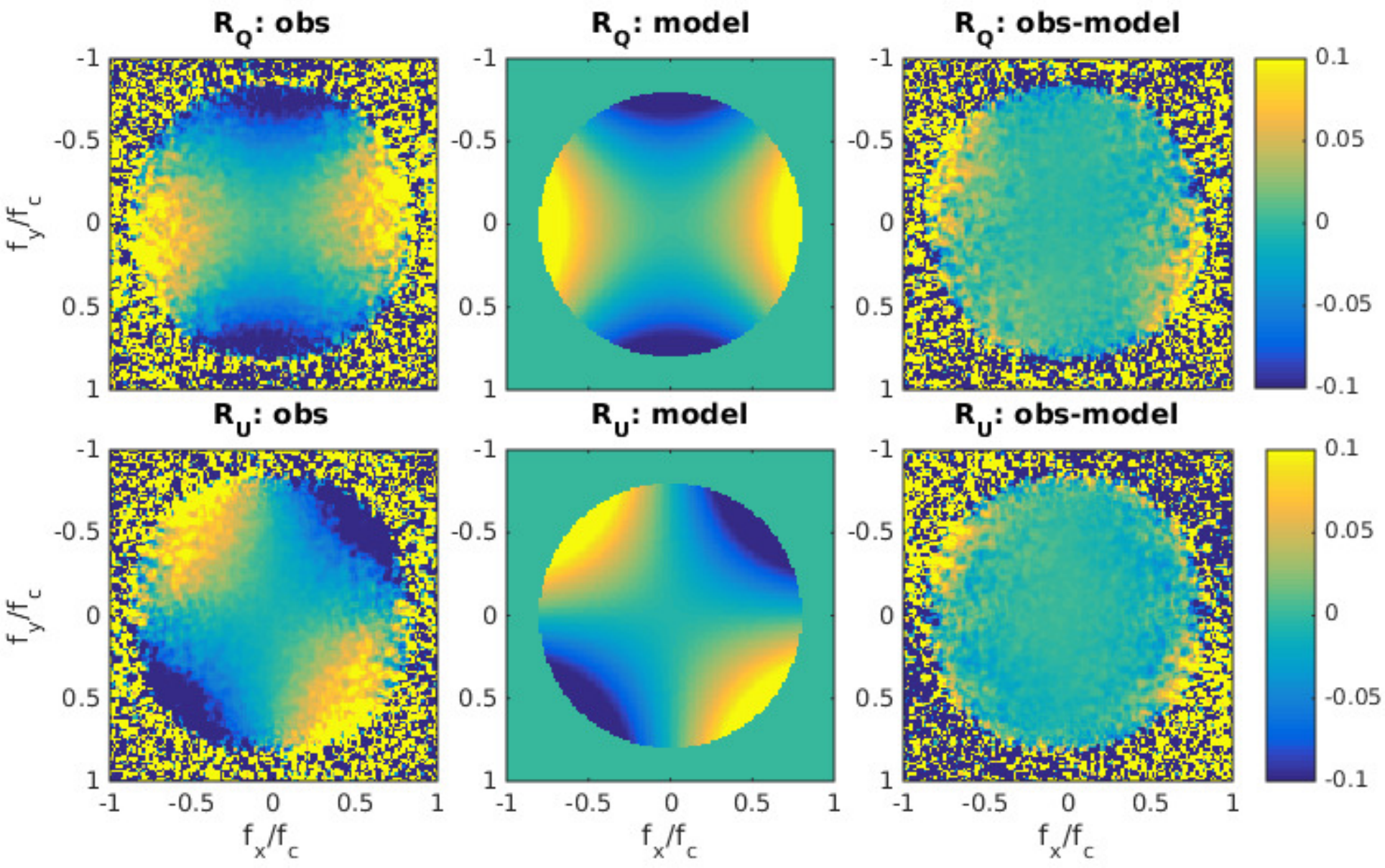}
\end{center}
\caption{
\ru{
Наблюдаемый и модельный модуль $\mathcal{R}$, а также их разность. Верхняя строчка --- параметр Стокса $Q$, нижняя --- $U$. По осям пространственная частота, нормированная на частоту среза. Диапазон отображаемых яркостей для всех панелей одинаковый, приведен справа. Наблюдение 11 декабря 2019 года в полосе 550~нм. Моделирование описано в разделе \ref{sec:parmodel}.}
\en{Observational and modelled absolute values of $\mathcal{R}$, and their difference. Upper row --- Stokes $Q$, lower --- Stokes $U$. Spatial frequency normalized by cut--off frequency is along the axes. The displayed range of signal for all the panels is the same and is given at the right side. The date of observation is 2019-12-11, the band is $550$~nm. Modelling is described in section~\ref{sec:parmodel}.}
\label{fig:parmodel}}
\end{figure}

\ru{\section{Результаты}}
\en{\section{Results}}

\ru{\subsection{Яркость оболочки}}
\en{\subsection{Envelope brightness}}
\label{sec:parmodel}

\ru{Первая особенность околозвездной оболочки Бетельгейзе, которую мы обсудим --- это вариации ее полной яркости относительно звезды. Из Рис.~\ref{fig:images} видно, что яркость оболочки возрастает с уменьшением длины волны, а также она больше когда полный поток от объекта меньше (в минимуме). Несмотря на наглядность изображений, их прямая интерпретация затруднена сглаживанием с ФРТ. Предпочитителен анализ непосредственно величины $\mathcal{R}$, оценка которой несмещенная.}
\en{We will start the discussion by considering the total brightness of the resolved envelope. From Fig.~\ref{fig:images} one can immediately see that the relative brightness of the envelope is larger for shorter wavelenghts and when the total flux from the object is less (in the minimum). However, quantitative interpretation of images is complicated by smoothing by the point spread function. A direct consideration of $\mathcal{R}$ is more preferable, as long as the latter is an unbiased estimator \citep{Safonov2019a}.}

\begin{figure}[t!]
\centering
\begin{tabular}{c}
\includegraphics[width=160mm]{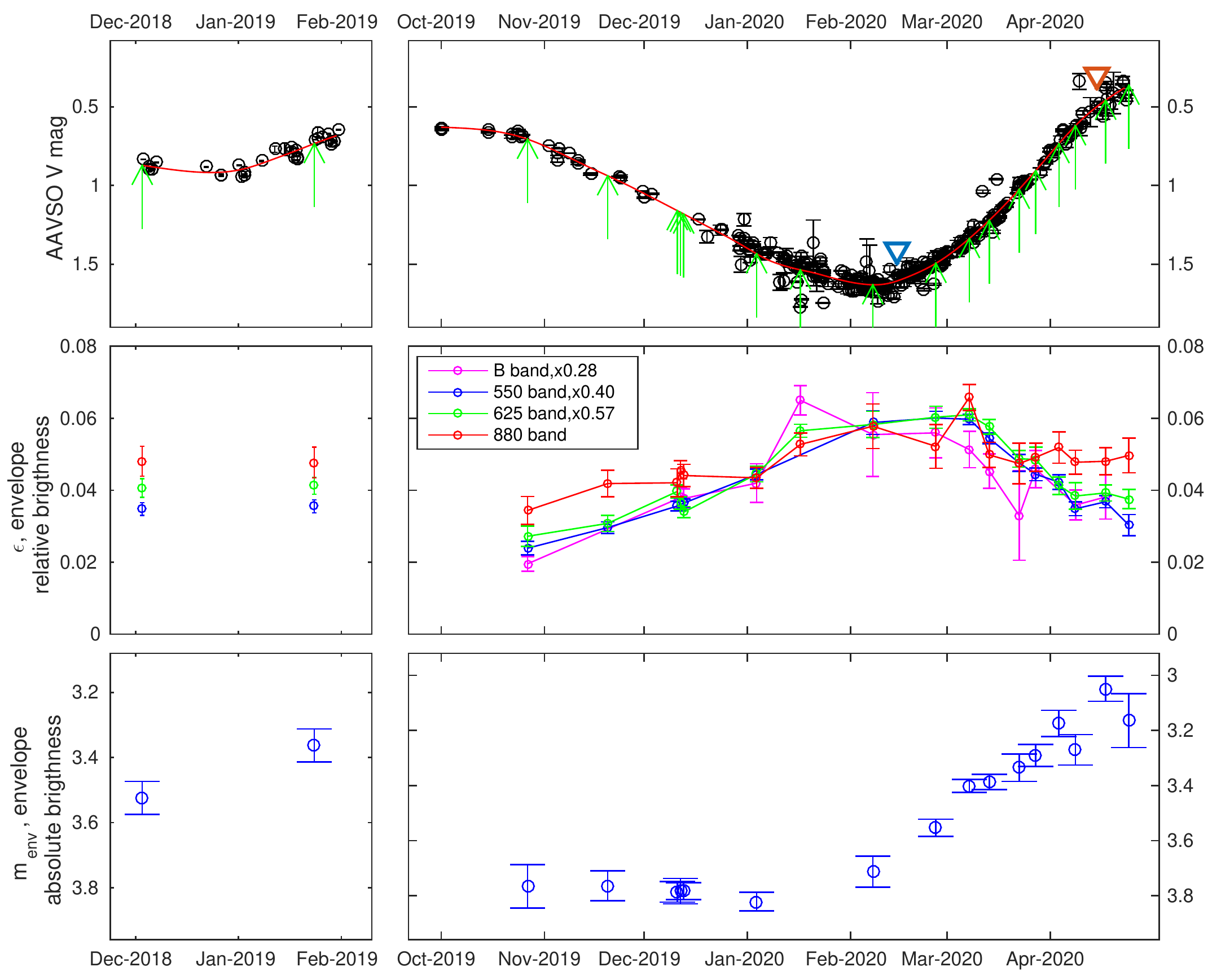} \\
\end{tabular}
\caption{\ru{Верхний ряд: черные кружки --- индивидуальные оценки блеска в полосе $V$ по данным AAVSO, красная линия --- сглаженная кривая блеска. Зеленые стрелки отмечают моменты времени, когда были выполнены наблюдения методом ДСП. Синий и оранжевый треугольники отмечают моменты спектральных наблюдений \citet{Levesque2020} и наши, соответственно. Средний ряд: величина $\epsilon$, харектеризующая амплитуду butterfly pattern в модуле $\mathcal{R}$. Она может рассматриваться как proxy поляризованной яркости оболочки относительно полной яркости объекта.
Нижний ряд: величина $m_\mathrm{env}$ с точностью до постоянной аддитивной добавки характеризующая абсолютную яркость оболочки в поляризованном потоке.}
\en{Upper row: black circles are for individual AAVSO $V$ magnitude estimations. Red line is a smoothed light curve. Green arrows denote moments of DSP observations. Blue and orange triangles denote the moments of \citet{Levesque2020} and our spectral observations, respectively. Middle row: the $\epsilon$ parameter which characterizes the amplitude of butterfly pattern in absolute value of $\mathcal{R}$. It is a proxy of polarized brightness of the envelope relative to the total brightness of the object. Lower row: $m_\mathrm{env}$ characterizes the absolute polarized brightness of the envelope with some constant addition.}
\label{fig:lc}}
\end{figure}

\ru{В предположении того что угловой размер рассеивающей оболочки постоянен, амплитуда butterfly pattern в модуле $\mathcal{R}$ будет пропорциональна ее яркости относительно звезды. Мы выполнили оценку последней путем аппроксимации измерений $\mathcal{R}$ следующими выражениями:}
\en{Under an assumption that the angular size of the scattering envelope is constant, the amplitude of the butterfly pattern in the absolute value of $\mathcal{R}$ will be proportional to the envelope brightness relative to the stellar one. We estimated this amplitude by approximation of $\mathcal{R}$ measurements using the following expressions:}
\begin{equation}
|\mathcal{R}_Q(\boldsymbol{f})| = 1 + q - \epsilon \frac{|f|^2}{f_c^2} \cos 2\theta,
\label{eq:RmodelQ}
\end{equation}
\begin{equation}
|\mathcal{R}_U(\boldsymbol{f})| = 1 + u - \epsilon \frac{|f|^2}{f_c^2} \sin 2\theta,
\label{eq:RmodelU}
\end{equation}
\ru{где $|f|$ и $\theta$ --- полярные координаты в пространстве частот. $q, u, \epsilon$ --- параметры модели, первые два характеризуют сдвиг $\mathcal{R}$, возникающий вследствие ненулевой полной поляризации объекта, а третий дает оценку амплитуды butterfly pattern. $f_c$ --- частота среза. Для возможности прямого сравнения оценок при аппроксимации во всех четырех полосах мы взяли частоту среза $f_c=2.84\times10^6$~RAD$^{-1}$, соответствующую $\lambda=880$~нм и диаметру апертуры 2.5~м. Мы подчеркиваем, что параметр $\epsilon$ лишь пропорционален относительной яркости оболочки, и то при условии постоянства ее угловых размеров.}
\en{where $|f|$ and $\theta$ are polar coordinates in the Fourier space. $q, u, \epsilon$ are the model parameters, the first two characterize the shift of $\mathcal{R}$ due to non--zero total polarization of the object. $\epsilon$ is the estimate of the amplitude of butterfly pattern. $f_c$ is the cut--off frequency of the optical system. In order to compare directly the estimations for all the bands we took the same cut--off frequency $f_c=2.84\times10^6$~RAD$^{-1}$ for them, which corresponds to $\lambda=880$~nm and an aperture diameter of 2.5~m. We emphasize that $\epsilon$ is only proportional to the relative brightness of the envelope.}

\ru{
Аппроксимация законом (\ref{eq:RmodelQ},\ref{eq:RmodelU}) выполнялась в области частот от 0.1$f_c$ до 0.4$f_c$, при этом применялось взвешивание на обратный квадрат ошибки величины $\mathcal{R}$ (метод ее оценки приведен в \cite{Safonov2019a}). Результаты для наблюдения 11 декабря 2019 года в полосе 550~нм приведены на Рис.~\ref{fig:parmodel}, наблюдается неплохое согласие. 

Зависимость $\epsilon$ от времени во всех полосах приведена на Рис.~\ref{fig:lc}, в среднем ряду. Она подтверждает качественные выводы из Рис.~\ref{fig:images}. Во-первых, яркость оболочки значительно больше в коротковолновых полосах. Так, в полосе $B$ оболочка ярче в 3.6 раз, чем в полосе $880$. Во-вторых, яркость оболочки относительно звезды возрастает в минимуме блеска.}

\en{
The approximation of observational data by the law (\ref{eq:RmodelQ},\ref{eq:RmodelU}) was performed at the frequencies between 0.1$f_c$ and 0.4$f_c$. Weighting by the inverse squared error of $\mathcal{R}$ was applied as well. The algorithm for estimating the $\mathcal{R}$ error in DSP method is provided by \citet{Safonov2019a}. The approximation is illustrated for one particular observation by Fig.~\ref{fig:parmodel}, very good agreement can be seen.

The dependence of $\epsilon$ on time is given for all the bands in middle row of Fig.~\ref{fig:lc}. It confirms the qualitative conclusions from Fig.~\ref{fig:images}. The envelope is much brighter at shorter wavelenghts: $\epsilon$ is 3.6 times brighter in $B$ band than at $880$~nm. On the other hand, the relative brightness of the envelope is larger in the minimum.}


\ru{Интересно проверить, постоянна ли абсолютная яркость оболочки на фоне наблюдаемых вариаций яркости звезды. Для этого мы вычислили следующую величину:}
\en{It is interesting to consider how the absolute brightness of the envelope behaves against the background of stellar variability. For that we computed the following quantity:}
\begin{equation}
m_\mathrm{env} = -2.5\log_{10}\epsilon + m_\star,
\end{equation}
\ru{где $m_\star$ --- звездная величина звезды на момент наблюдения (получено интерполяцией наблюдений AAVSO в полосе $V$). Эта процедура была выполнена только для полосы $550$, центральная длина волны которой весьма близка к аналогичной величине для полосы $V$. $m_\mathrm{env}$ с точностью до постоянной добавки характеризует поляризованную составляющую абсолютной яркости оболочки.}
\en{where $m_\star$ is the magnitude of the star (obtained by the interpolation of the AAVSO $V$ magnitude data). This procedure was done only for the 550 band, because its central wavelength is close to that of the $V$ band. $m_\mathrm{env}$ corresponds to the polarized absolute brightness of the envelope with some constant addition.}
\ru{
Оценки $m_\mathrm{env}$ представлены на Рис.~\ref{fig:lc}, в нижнем ряду. В период с конца октября по конец января эта величина была постоянна, в то время как звезда ослабевала. Это подтверждает выводы \cite{Levesque2020} о том что звезда ослабевает вследствие увеличение околозвездного поглощения на направлении к наблюдателю. В тоже время освещенность оболочки, окружающей звезду, не изменилась и ее яркость сохранялась постоянной.

В конце января --- начале февраля, незадолго до разворота на кривой блеска, ситуация изменилась, яркость оболочки начала интенсивно возрастать. К началу апреля оболочка стала в 1.7 раза ярче чем была в октябре 2019 --- январе 2020 и в $1.2-1.3$ раза ярче, чем за год до этого. 
}

\en{
The resulting $m_\mathrm{env}$ is presented in the lower row of Fig.~\ref{fig:lc}. During the period from October 2019 to January 2020 this quantity was constant while the star became significantly fainter. This supports the conclusions of \citet{Levesque2020} that the star decreased its brightness due to a dust cloud localized on the line of sight. The envelope's brightness stayed constant because the illumination conditions of the envelope surrounding the star did not change.

Since February 2020 the situation had changed, the envelope's brightness started to rise dramatically. By the end of April 2020 it became 2.1 times brighter than in October 2019 -- January 2020 and 1.5 times brighter than a year before.
}

\begin{figure*}
\begin{center}
\includegraphics[scale=0.3]{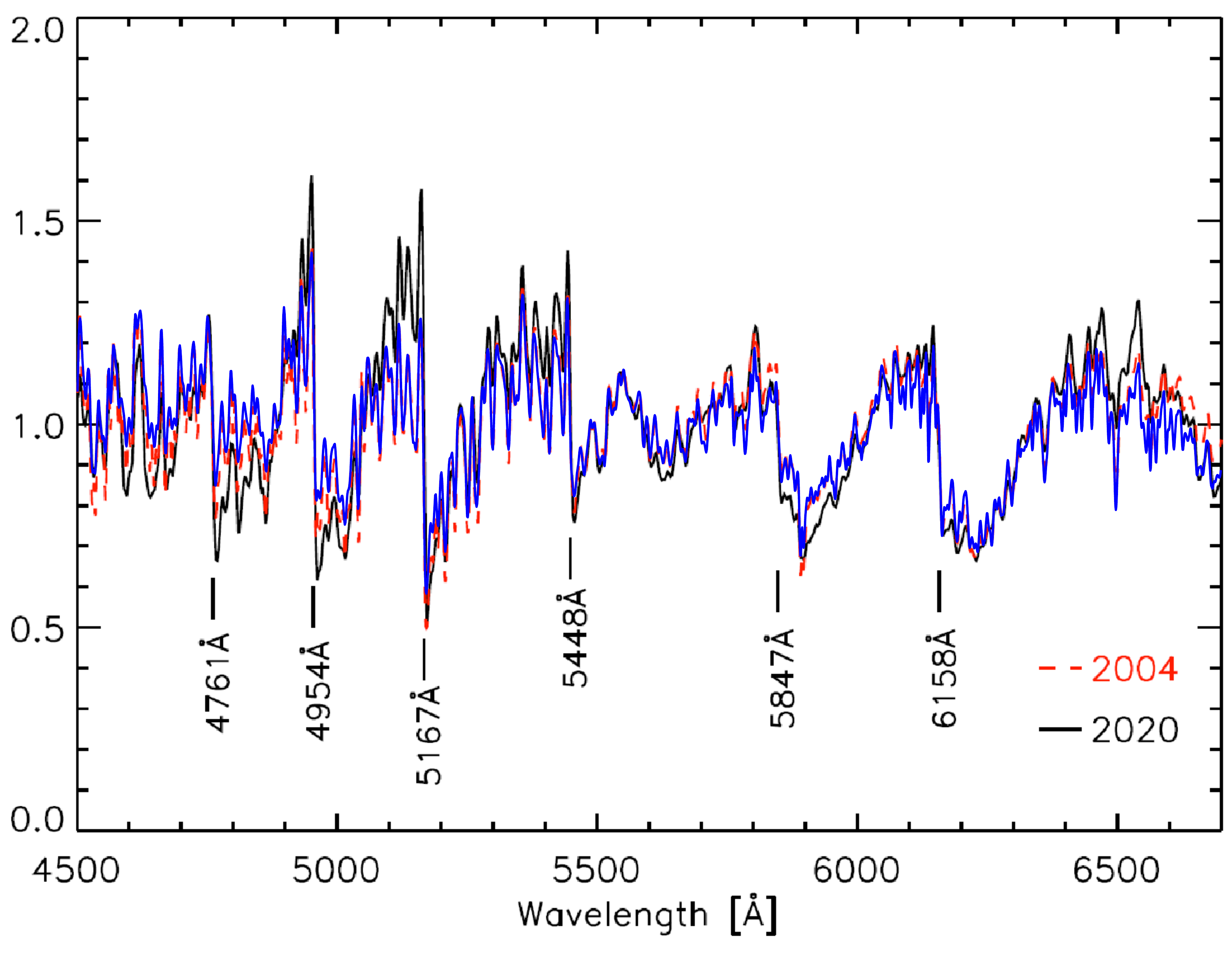}
\end{center}
\caption{\ru{
Участок спектра Бетельгейзе, рис. 2 из работы \citet{Levesque2020}. Наш спектр  полученный в 6 апреля 2020 года обозначен синим.}
\en{
Part of the Betelgeuse spectrum (Fig.~2 from \citep{Levesque2020}). Our spectrum secured on 6th April 2020 is overplotted as blue line. 
}
}
\label{fig:TDS}
\end{figure*}

\ru{В тоже время полученный нами 6 апреля 2020 года спектр хорошо согласуется со спектром из \cite{Levesque2020}, полученным в 2004 году, т.е. до минимума, см. Рис.~\ref{fig:TDS}. Изменение температуры звезды на протяжении всего минимума не было значительным и не может объяснять вариации яркости звезды (см. обоснование в \cite{Levesque2020}) и оболочки.}

\en{At the same time the spectrum, which we obtained on April, 6th 2020, almost coincides with the spectra from 2004 and 2020 presented by \citep{Levesque2020}, see Fig.~\ref{fig:TDS}. The change in the temperature of the photosphere during the whole minimum was not significant and cannot explain the variability of the star and its envelope, see detailed discussion in \citep{Levesque2020}.}




\ru{\subsection{Морфология оболочки}}
\en{\subsection{Morphology of the envelope}}

\ru{
Из Рис.\ref{fig:images} можно видеть, что оболочка не обладает центральной симметрией. Относительные вариации поляризованного потока на разных позиционных углах составляют $2-4$. Морфология оболочки меняется как в зависимости от длины волны, так и времени. Так, южная половина выглядит ослабленной на входе звезды в минимум блеска. Об этом также сообщают \citet{Montarges2020}. С конца февраля 2020 года распределение яркости в оболочке становится более симметричным. К апрелю южная часть звезды становится напротив ярче, чем оболочка в среднем. Обращает на себя внимание также северо--восточное пятно, которое возникло в конце ноября, достигло максимальной яркости в начале февраля и к апрелю практически исчезло.
}
\en{ 
From Fig.\ref{fig:images} it can be seen that the scattering envelope significantly deviates from point symmetry. The relative variation of polarized flux at different position angles is around $2-4$. The envelope morphology changes with wavelength and time. The southern part of the envelope appears fainter in the first half of the minimum --- the fact which was reported by \citet{Montarges2020}. Since the mid February 2020 southern part had been brightening and by the end of April it dominated the envelope, especially in 625 and 880~nm bands. One can also note a North-Eastern feature which emerged in November 2019, reached maximum brightness in the mid February and almost disappeared in April 2020.
}

\ru{
Изменения морфологии наиболее удобно прослеживаются с помощью анимации, которую мы собрали из изображений на $\lambda=550$~нм. В этой анимации изображения были попиксельно интерполированы на равномерную сетку по времени. Также отсчет в каждом пикселе был умножен на величину $2.512^{-m_V}$, где $m_V$ --- звездная величина AAVSO, интерполированная на момент наблюдения. Таким образом на анимации приведена абсолютная поляризованная яркость оболочки. В этом ее отличие от рисунка \ref{fig:images}, где дана поляризованная яркость относительно звезды. Представление в абсолютной поляризованной яркости позволяет еще раз убедиться в том что до момента наименьшей яркости звезды яркость оболочки была постоянна, а после начала интенсивно расти.
}

\en{
In order to make the changes in morphology more visible we compiled an animation from the images at 550~nm, which can be found via link\footnote{\url{http://lnfm1.sai.msu.ru/kgo/mfc_Betelgeuse_en.php}}. In that animation the images were interpolated pixel--wise on a equidistant temporal grid. Besides, the value in each pixel was multiplied by $2.512^{-m_V}$, where $m_V$ is an AAVSO magnitude interpolated on the moment of observation. Therefore in the animation an absolute polarized brightness is presented. This is in contrast with Fig.\ref{fig:images} where we give polarized brightness relative to the star. The absolute polarized brightness allows one to see once more that before the bottom of the minimum of the star the brightness of the envelope was constant, however it started to rise after this moment. 
}

\ru{
Описанные структуры видимы на длинах волн 465, 550 и 625~нм. В полосе 880~нм они проявлены значительно меньше. Однако изображения для наблюдений в фокусе Нэсмита, выполненных в этой полосе, могут быть искажены эффектами инструментальной поляризации. Тем не менее, полная яркость оболочки, о которой речь шла в предыдущем подразделе, оценивается вполне надежно. Изображения, восстановленные по наблюдениям в фокусе Кассегрена (11 и 12 декабря 2019 года), вполне надежны во всех полосах.
}

\en{
These structures are visible very clearly at wavelenghts of 465, 550, 625~nm. At 880~nm they are less contrast. However, observations taken in the Nasmyth focus using this band can be distorted by the effects of instrumental polarization. Nevertheless, net polarized brightness of the envelope, which was discussed in previous subsection, is estimated quite reliably. The images restored using observations in the Cassegrain focus (11th and 12th December 2019) are accurate in all the bands.
}

\ru{
Обнаруженные детали морфологии оболочки являются следствием неоднородности плотности основания пылевого ветра, либо неоднородностями его освещения звездой. Оба эффекта могут быть порождены крупными конвективными ячейками на поверхности звезды. Неоднородности околозвездной оболочки Бетельгейзе на расстояниях $1-3R_\star$ от фотосферы были обнаружены ранее \citet{Haubois2019,Kervella2016,OGorman2017}. 
}

\en{
The features of Betelgeuse's envelope can be induced by inhomogenuities of the density at the base of dusty wind or by inhomogenuities of illumination of this wind by the photosphere. Both effects are likely to originate from large convective cells on the stellar surface. Similar features of Betelgeuse's envelope were detected at distances of $1-3R_\star$ from the photosphere before by a number \citet{Haubois2019,Kervella2016,OGorman2017}.
}

\ru{
\section{Заключение}
}

\en{
\section{Conclusion}
}

\ru{
Мы выполнили наблюдения компактной оболочки Бетельгейзе, обнаруженной ранее на VLT/NaCO \cite{Haubois2019} и VLT/SPHERE \cite{Montarges2020}. Эта оболочка располагается на расстояниях порядка 1 радиуса звезды от фотосферы и вероятно является основанием пылевого ветра. Она показывает значительные неоднородности яркости. Наши наблюдения, выполненные в 17 эпох, распределенных по интервалу длительностью полгода, показывают изменения неоднородностей во времени. Картина неоднородностей значительно скоррелирована для последовательных наблюдений. Характерные времена, на которых развиваются неоднородности, составляют 2--3 месяца.}

\en{
We present observations of the innermost part of Betelgeuse's envelope found previously at VLT/NaCO by \citep{Haubois2019} and at VLT/SPHERE by \citep{Haubois2019}. The envelope is located at $\approx1R_\star$ above the photosphere, it is likely to be the base of dusty wind. The envelope was found to be highly inhomogeneous. Our observations on 17 dates spread over a half the year allowed us to trace the changes of the envelope morphology. For consecutive epochs the observations are highly correlated. A typical timescale of envelope features variability amounts to 2--3 months.
}

\ru{
Наши наблюдения охватывают период глубокого минимума звезды c октября 2019 года по апрель 2020 года. В течение этого минимума звезда ослабла на $1^m$ за $\approx3.5$~месяца. Спектральные наблюдения \citet{Levesque2020} показывают что ослабление объясняется увеличением количества пыли в атмосфере, а не уменьшением температуры фотосферы. По нашим данным полная поляризованная яркость оболочки была постоянна на входе в минимум. Это говорит о том что увеличение околозвездного поглощения было значимо только в направлении на наблюдателя, в том время как освещенность оболочки на других направления не изменилась.}
\en{
Our observations cover the period of a deep minimum of Betelgeuse's brightness from October 2019 to April 2020. During this minimum the star became fainter by $1^m$ over the period of 3.5~months. Spectral observations by \citet{Levesque2020} demonstrate that the dimming can be explained by an increase in the amount of absorbing dust in the stellar atmosphere, rather than a decrease in the temperature of the photosphere. According to our data, the total polarized brightness of the envelope was constant during the phase of fainting. This suggests that the increase in circumstellar absorption was significant only for the directions close to the line of sight, while the illumination of the envelope in other directions did not change.}

\ru{
Восстановление яркости Бетельгейзе было примерно в 2 раза более быстрым, чем ослабление. Наши спектральные наблюдения 6 апреля 2020 года показали, что подъем яркости также не был связан с изменениями температуры фотосферы. Во время подъема яркости звезды поведение оболочки изменилось: ее поляризованная яркость начала возрастать и к 24 апрелю она возросла уже в 2.1 раза. Интересно, что в течение 2 месяцев до наступления минимума южная половина оболочки была гораздо темнее чем северная. На выходе из минимума ситуация поменялась на противоположную: южная половина доминировала.
}
\en{
The rise of Betelgeuse's brightness was approximately two times faster than the fall. Our spectral observations on 2020-04-06 show that the brightening was not related to the change in photosphere temperature either. During the recovery from the minimum the behaviour of the envelope changed dramatically: its polarized brightness started to rise and by 2020-04-24 it was 2.1 times brighter than at the star's minimum. It is noteworthy that, within 2 months before the minimum, the southern half of the envelope was much darker than the northern (see also \citep{Montarges2020}). During the recovery from the minimum, the situation changed to the opposite: the southern half dominated the envelope.
}

\ru{
Возрастание яркости околозвездной оболочки может быть вызвано уменьшением самопоглощения рассеянного света в ней. При входе в минимум и в минимуме оболочка ближе к звезде, самопоглощение в ней велико. На выходе из минимума оболочка расширяется и оптическая толщина в ней падает, в результате чего она рассеивает больше света. Вероятно также что общее количество рассеивающего вещества в оболочке возросло после минимума.}

\en{
The rise of the envelope brightness could be caused by a reduced self--absorption of scattered radiation in it. At the beginning of the minimum the cloud was closer to the star, self--absorption was larger. Then the cloud moved away from the star and its optical depth decreased, as a result, it scattered more light. It is likely that the total amount of scattering material in the envelope increased after the minimum.
}

\ru{
Глубокое затмение Бетельгейзе дает возможность изучить динамику пылевого облака в атмосфере красного сверхгиганта. Для построения количественной картины этого процесса важно задействовать не только наблюдения в рассеянном свете, подобные нашим и \citet{Montarges2020}, но и наблюдения в ИК-диапазоне, которые позволили бы зарегистрировать тепловое излучение образовавшейся пыли. Если звезда действительно потускнела из-за образования пылевого облака, то уменьшение ее яркости должно быть вызвано не только рассеянием, но и поглощением со стороны пыли. А поскольку пылинки, поглощая свет звезды, нагреваются, мы могли бы ожидать увеличения уровня инфракрасного излучения. 
}

\en{
The deep minimum of Betelgeuse provides an opportunity to follow the dynamics of dusty clouds in the atmosphere of red supergiant star. In order to construct a quantitative model of this process it is necessary to employ not only scattered light observations similar to ours and that of \citet{Montarges2020}, but IR observations of  thermal radiation of newly formed dust as well. If the star really faded due to the formation of a dust cloud, then a decrease in its brightness should be caused not only by scattering, but also by absorption by the dust. Since the dust particles, absorbing the stellar radiation, heat up, we could expect an increase in the level of IR radiation.
}

\ru{
Наличие фотометрических данных или спектров звезды в ближнем и среднем ИК-диапазонах в моменты времени до минимума и во время минимума блеска может позволить увидеть возрастание ИК-излучения в момент затмения и благодаря этому подтвердить гипотезу об образовании пылевого облака. Кроме того, это дало бы возможность оценить оптическую толщу и массу образовавшегося пылевого вещества. Используя спектральные или фотометрические данные в ИК-диапазоне, а также имеющиеся наблюдения рассеянного излучения в видимом диапазоне до минимума, во время минимума и после выхода из минимума блеска, мы могли бы проследить эволюцию пылевого облака от момента образования.
}

\en{
Photometric or spectral data for Betelgeuse in near--IR and mid--IR for the moments before the minimum and close to the bottom of the minimum could made it possible to detect an increase of IR radiation associated with the dimming and confirm the model of dusty cloud formation. Besides, it would allow us to estimate the optical depth and total mass of the condensed dust. If we had at our disposal IR photometric or spectral data in addition to available resolved scattered light observations in visual range, we would be able to trace the evolution of the dusty cloud from the moment of its origin.
}

\ru{
Наша работа показывает как  сравнительно простой прибор реализующий дифференциальную спекл-поляриметрию может использовать время большого телескопа для получения новой информации о звёздах на поздних стадиях эволюции с  высоким разрешением. Полезные данные могут быть получены в плохих атмосферных условиях, в сумерки и даже днём. Такие наблюдения, выполняемые регулярно, могли бы быть использованы в качестве материала для проверки и уточнения моделей звёздного ветра старых звёзд, подобных разрабатываемым \citet{Freytag2017}. 
}

\en{
From methodological point of view our work demonstrates that a relatively simple instrument mounted at a large telescope could take advantage of the mediocre atmospheric conditions, twilight time or even daytime to study envelopes of evolved stars at spatial scales comparable to 1 stellar radius. Such observations of multiple objects being conducted on a regular basis would be useful for constraining the models of stellar winds similar to \citep{Freytag2017}.
}

\ru{
\section*{Благодарности}

Мы благодарим Татарникова А.М. за помощь в проведении наблюдений. Спекл-поляриметр создан при поддержке программы развития МГУ.
}

\en{
\section*{Acknowledgements}

We are grateful to Tatarnikov A. M. for the help with observations. The development and construction of the speckle polarimeter of the 2.5-m telescope has been funded by the M. V. Lomonosov Moscow State University Program of Development.
}


\bibliography{biblio}

\begin{thebibliography}{17}
\providecommand{\natexlab}[1]{#1}
\providecommand{\url}[1]{\texttt{#1}}
\expandafter\ifx\csname urlstyle\endcsname\relax
  \providecommand{\doi}[1]{doi: #1}\else
  \providecommand{\doi}{doi: \begingroup \urlstyle{rm}\Url}\fi

\bibitem[{Clarke}(2010)]{Clarke2010}
D.~{Clarke}.
\newblock \emph{{Stellar Polarimetry}}.
\newblock 2010.

\bibitem[Cotton et~al.(2020)Cotton, Bailey, Horta, Norris, and
  Lomax]{Cotton2020}
D.~V. Cotton, J.~Bailey, A.~D. Horta, B.~R.~M. Norris, and J.~R. Lomax.
\newblock Multi-band aperture polarimetry of betelgeuse during the
  2019{\textendash}20 dimming.
\newblock \emph{Research Notes of the {AAS}}, 4\penalty0 (3):\penalty0 39, mar
  2020.
\newblock \doi{10.3847/2515-5172/ab7f2f}.
\newblock URL \url{https://doi.org/10.3847%2F2515-5172%2Fab7f2f}.

\bibitem[{Fedotyeva} et~al.(2020){Fedotyeva}, {Tatarnikov}, {Safonov},
  {Shenavrin}, and {Komissarova}]{Fedotyeva2020}
A.~A. {Fedotyeva}, A.~M. {Tatarnikov}, B.~S. {Safonov}, V.~I. {Shenavrin}, and
  G.~{Komissarova}.
\newblock {A Model of the Dust Envelope of the Carbon Mira V CrB from
  Photometry, Infrared Spectroscopy, and Speckle Polarimetry}.
\newblock \emph{Astronomy Letters}, 46\penalty0 (1):\penalty0 41--60, Jan.
  2020.
\newblock \doi{10.1134/S1063773720010016}.

\bibitem[{Freytag} et~al.(2017){Freytag}, {Liljegren}, and
  {H{\"o}fner}]{Freytag2017}
B.~{Freytag}, S.~{Liljegren}, and S.~{H{\"o}fner}.
\newblock {Global 3D radiation-hydrodynamics models of AGB stars. Effects of
  convection and radial pulsations on atmospheric structures}.
\newblock \emph{\aap}, 600:\penalty0 A137, Apr. 2017.
\newblock \doi{10.1051/0004-6361/201629594}.

\bibitem[{Guinan} et~al.(2020){Guinan}, {Wasatonic}, {Calderwood}, and
  {Carona}]{Guinan2020}
E.~{Guinan}, R.~{Wasatonic}, T.~{Calderwood}, and D.~{Carona}.
\newblock {The Fall and Rise in Brightness of Betelgeuse}.
\newblock \emph{The Astronomer's Telegram}, 13512:\penalty0 1, Feb. 2020.

\bibitem[{Haubois} et~al.(2019){Haubois}, {Norris}, {Tuthill}, {Pinte},
  {Kervella}, {Girard}, {Kostogryz}, {Berdyugina}, {Perrin}, {Lacour},
  {Chiavassa}, and {Ridgway}]{Haubois2019}
X.~{Haubois}, B.~{Norris}, P.~G. {Tuthill}, C.~{Pinte}, P.~{Kervella}, J.~H.
  {Girard}, N.~M. {Kostogryz}, S.~V. {Berdyugina}, G.~{Perrin}, S.~{Lacour},
  A.~{Chiavassa}, and S.~T. {Ridgway}.
\newblock {The inner dust shell of Betelgeuse detected by polarimetric
  aperture-masking interferometry}.
\newblock \emph{\aap}, 628:\penalty0 A101, Aug. 2019.
\newblock \doi{10.1051/0004-6361/201833258}.

\bibitem[{Kervella} et~al.(2016){Kervella}, {Lagadec}, {Montarg{\`e}s},
  {Ridgway}, {Chiavassa}, {Haubois}, {Schmid}, {Langlois}, {Gallenne}, and
  {Perrin}]{Kervella2016}
P.~{Kervella}, E.~{Lagadec}, M.~{Montarg{\`e}s}, S.~T. {Ridgway},
  A.~{Chiavassa}, X.~{Haubois}, H.~M. {Schmid}, M.~{Langlois}, A.~{Gallenne},
  and G.~{Perrin}.
\newblock {The close circumstellar environment of Betelgeuse. III.
  SPHERE/ZIMPOL imaging polarimetry in the visible}.
\newblock \emph{\aap}, 585:\penalty0 A28, Jan. 2016.
\newblock \doi{10.1051/0004-6361/201527134}.

\bibitem[{Kornilov} et~al.(2014){Kornilov}, {Safonov}, {Kornilov}, {Shatsky},
  {Voziakova}, {Potanin}, {Gorbunov}, {Senik}, and {Cheryasov}]{Kornilov2014}
V.~{Kornilov}, B.~{Safonov}, M.~{Kornilov}, N.~{Shatsky}, O.~{Voziakova},
  S.~{Potanin}, I.~{Gorbunov}, V.~{Senik}, and D.~{Cheryasov}.
\newblock {Study on Atmospheric Optical Turbulence above Mount Shatdzhatmaz in
  2007-2013}.
\newblock \emph{\pasp}, 126:\penalty0 482--495, May 2014.
\newblock \doi{10.1086/676648}.

\bibitem[{Levesque} and {Massey}(2020)]{Levesque2020}
E.~M. {Levesque} and P.~{Massey}.
\newblock {Betelgeuse Just Is Not That Cool: Effective Temperature Alone Cannot
  Explain the Recent Dimming of Betelgeuse}.
\newblock \emph{\apjl}, 891\penalty0 (2):\penalty0 L37, Mar. 2020.
\newblock \doi{10.3847/2041-8213/ab7935}.

\bibitem[{Michelson} and {Pease}(1921)]{Michelson1921}
A.~A. {Michelson} and F.~G. {Pease}.
\newblock {Measurement of the Diameter of {\ensuremath{\alpha}} Orionis with
  the Interferometer.}
\newblock \emph{\apj}, 53:\penalty0 249--259, May 1921.
\newblock \doi{10.1086/142603}.

\bibitem[{Montarges et al}(2020)]{Montarges2020}
M.~{Montarges et al}.
\newblock {ESO Telescope Sees Surface of Dim Betelgeuse}.
\newblock \emph{ESO press release}, Feb. 2020.

\bibitem[{Natta} and {Panagia}(1984)]{Natta1984}
A.~{Natta} and N.~{Panagia}.
\newblock {Extinction in inhomogeneous clouds.}
\newblock \emph{\apj}, 287:\penalty0 228--237, Dec. 1984.
\newblock \doi{10.1086/162681}.

\bibitem[{Norris} et~al.(2012){Norris}, {Tuthill}, {Ireland}, {Lacour},
  {Zijlstra}, {Lykou}, {Evans}, {Stewart}, and {Bedding}]{Norris2012}
B.~R.~M. {Norris}, P.~G. {Tuthill}, M.~J. {Ireland}, S.~{Lacour}, A.~A.
  {Zijlstra}, F.~{Lykou}, T.~M. {Evans}, P.~{Stewart}, and T.~R. {Bedding}.
\newblock {A close halo of large transparent grains around extreme red giant
  stars}.
\newblock \emph{\nat}, 484:\penalty0 220--222, Apr. 2012.
\newblock \doi{10.1038/nature10935}.

\bibitem[{O'Gorman} et~al.(2017){O'Gorman}, {Kervella}, {Harper}, {Richards},
  {Decin}, {Montarg{\`e}s}, and {McDonald}]{OGorman2017}
E.~{O'Gorman}, P.~{Kervella}, G.~M. {Harper}, A.~M.~S. {Richards}, L.~{Decin},
  M.~{Montarg{\`e}s}, and I.~{McDonald}.
\newblock {The inhomogeneous submillimeter atmosphere of Betelgeuse}.
\newblock \emph{\aap}, 602:\penalty0 L10, June 2017.
\newblock \doi{10.1051/0004-6361/201731171}.

\bibitem[{Safonov} et~al.(2019{\natexlab{a}}){Safonov}, {Lysenko},
  {Goliguzova}, and {Cheryasov}]{Safonov2019a}
B.~{Safonov}, P.~{Lysenko}, M.~{Goliguzova}, and D.~{Cheryasov}.
\newblock {Differential speckle polarimetry at Cassegrain and Nasmyth foci}.
\newblock \emph{\mnras}, 484:\penalty0 5129--5141, Apr. 2019{\natexlab{a}}.
\newblock \doi{10.1093/mnras/stz288}.

\bibitem[{Safonov} et~al.(2017){Safonov}, {Lysenko}, and {Dodin}]{Safonov2017}
B.~S. {Safonov}, P.~A. {Lysenko}, and A.~V. {Dodin}.
\newblock {The speckle polarimeter of the 2.5-m telescope: Design and
  calibration}.
\newblock \emph{Astronomy Letters}, 43\penalty0 (5):\penalty0 344--364, May
  2017.
\newblock \doi{10.1134/S1063773717050036}.

\bibitem[{Safonov} et~al.(2019{\natexlab{b}}){Safonov}, {Dodin}, {Lamzin}, and
  {Rastorguev}]{Safonov2019b}
B.~S. {Safonov}, A.~V. {Dodin}, S.~A. {Lamzin}, and A.~S. {Rastorguev}.
\newblock {The Circumstellar Envelope of the Semiregular Variable Star V CVn}.
\newblock \emph{Astronomy Letters}, 45\penalty0 (7):\penalty0 453--461, July
  2019{\natexlab{b}}.
\newblock \doi{10.1134/S1063773719070065}.

\end{thebibliography}

\newpage

\appendix

\section{Observational log}
\label{app:obslog}

\ru{Наблюдения Бетельгейзе приведены в Табл.~\ref{table:obslog}. Для этих наблюдений по возможности выбирались плохие условия по прозрачности и качеству изображения. Наблюдения в марте--апреле также выполнялись в основном в вечерние сумерки либо непосредственно перед заходом Солнца, поэтому данные по качеству изображения, измеренному астроклиматическим монитором, отсутствуют.}
\en{
Observations of Betelgeuse are listed in Table~\ref{table:obslog}. We conducted them preferentially in bad conditions in terms of atmospheric turbulence and transparency. Observations in March--April 2020 were conducted mostly during evening twilight or just before the sunset. Because of that there is no concurrent seeing monitor measurements.
}

\begin{table}[h!]
\tabcolsep=0.15cm
\begin{small}
\begin{center}
\caption{\ru{Условия наблюдений Бетельгейзе. Колонки 1. фокальная станция: С --- Кассегрен, N --- Нэсмит; 2. Всемирное время; 3. Фотометрическая полоса; 4. Количество накопленных кадров; 5. Зенитное расстояние; 6. Высота Солнца; 7. Качество изображения по данным астроклиматического монитора \citep{Kornilov2014}.}
\en{Betelgeuse observations circumstances. Columns: 1. focal station: C --- Cassegrain, N --- Nasmyth; 2. Universal time; 3. Photometric band; 4. Number of accumulated frames; 5. Zenith distance; 6. Sun altitude; 7. Seeing according to automatic seeing monitor \citep{Kornilov2014}.}
\label{table:obslog}}
\begin{tabular}{cc}
\begin{tabular}[t]{ccccccc}
\hline
foc. & UT & band & $N_\mathrm{fr}$ & $z,^{\circ}$ & $h_\odot,^{\circ}$ & $\beta,^{\prime\prime}$ \\ 
\hline
N & 18-12-02 23:45 & 550 & 3007 & 41 & -49 & 0.97 \\
N & 18-12-02 23:48 & 625 & 3002 & 41 & -49 & 0.95 \\
N & 18-12-02 23:52 & 880 & 3002 & 42 & -48 & 0.99 \\
N & 19-01-23 20:21 & 550 & 3012 & 41 & -63 & 1.15 \\
N & 19-01-23 20:25 & 625 & 3011 & 42 & -63 & 1.18 \\
N & 19-01-23 20:28 & 880 & 3005 & 42 & -63 & 0.56 \\
N & 19-10-26 22:52 & B & 6002 & 44 & -50 & 0.68 \\
N & 19-10-26 22:41 & 550 & 3002 & 45 & -51 & 0.83 \\
N & 19-10-26 22:44 & 625 & 3005 & 45 & -51 & 0.82 \\
N & 19-10-26 22:48 & 880 & 3870 & 44 & -50 & 0.73 \\
N & 19-11-20 00:42 & B & 6003 & 41 & -37 & 1.20 \\
N & 19-11-20 00:33 & 550 & 3002 & 41 & -39 & 1.30 \\
N & 19-11-20 00:36 & 625 & 3004 & 41 & -38 & 1.35 \\
N & 19-11-20 00:38 & 880 & 3009 & 41 & -38 & 1.23 \\
N & 19-12-10 22:28 & 550 & 3509 & 37 & -63 & 1.06 \\
N & 19-12-10 22:22 & 625 & 3504 & 37 & -63 & 0.98 \\
N & 19-12-10 22:25 & 880 & 3507 & 37 & -63 & 1.40 \\
C & 19-12-11 21:38 & B & 6003 & 36 & -68 & 0.93 \\
C & 19-12-11 21:27 & 550 & 3005 & 36 & -69 & 0.95 \\
C & 19-12-11 21:30 & 625 & 3002 & 36 & -68 & 0.91 \\
C & 19-12-11 21:33 & 880 & 3014 & 36 & -68 & 0.97 \\
C & 19-12-12 20:58 & B & 6016 & 37 & -69 & 1.15 \\
C & 19-12-12 20:45 & 550 & 3008 & 38 & -69 & 1.02 \\
C & 19-12-12 20:48 & 625 & 3006 & 38 & -69 & 1.02 \\
C & 19-12-12 20:51 & 880 & 3002 & 38 & -69 & 0.96 \\
N & 20-01-03 18:40 & B & 6010 & 42 & -52 & 1.14 \\
N & 20-01-03 18:31 & 550 & 3011 & 43 & -51 & 1.11 \\
N & 20-01-03 18:33 & 625 & 3006 & 42 & -51 & 1.13 \\
N & 20-01-03 18:36 & 880 & 3003 & 42 & -52 & 1.13 \\
N & 20-01-16 22:35 & B & 6010 & 56 & -62 & 1.37 \\
N & 20-01-16 22:26 & 550 & 3005 & 54 & -63 & 1.41 \\
N & 20-01-16 22:29 & 625 & 3011 & 55 & -63 & 1.35 \\
N & 20-01-16 22:32 & 880 & 3002 & 55 & -63 & --- \\
N & 20-02-07 16:56 & B & 6006 & 39 & -27 & 1.54 \\
N & 20-02-07 16:46 & 550 & 3011 & 39 & -26 & 1.15 \\
N & 20-02-07 16:49 & 625 & 3002 & 39 & -26 & 1.40 \\
N & 20-02-07 16:52 & 880 & 3004 & 39 & -27 & 1.60 \\
\end{tabular}
&
\begin{tabular}[t]{ccccccc}
\hline
foc. & UT & band & $N_\mathrm{fr}$ & $z,^{\circ}$ & $h_\odot,^{\circ}$ & $\beta,^{\prime\prime}$ \\ 
\hline
N & 20-02-26 15:57 & B & 6008 & 38 & -12 & 1.50 \\  
N & 20-02-26 15:46 & 550 & 3005 & 38 & -11 & --- \\ 
N & 20-02-26 15:50 & 625 & 3012 & 38 & -11 & 1.54 \\
N & 20-02-26 15:53 & 880 & 3004 & 38 & -12 & 1.39 \\
N & 20-03-07 17:18 & B & 6002 & 40 & -25 & 1.47 \\  
N & 20-03-07 17:08 & 550 & 3002 & 39 & -23 & --- \\ 
N & 20-03-07 17:11 & 625 & 3002 & 39 & -23 & 1.04 \\
N & 20-03-07 17:14 & 880 & 3006 & 40 & -24 & 1.52 \\
N & 20-03-13 16:44 & B & 6003 & 39 & -17 & 0.96 \\  
N & 20-03-13 16:35 & 550 & 3007 & 38 & -16 & 1.14 \\
N & 20-03-13 16:37 & 625 & 3012 & 39 & -16 & 1.00 \\
N & 20-03-13 16:40 & 880 & 3004 & 39 & -17 & 1.00 \\
N & 20-03-22 16:45 & B & 6004 & 43 & -15 & --- \\   
N & 20-03-22 16:36 & 550 & 3008 & 42 & -14 & --- \\ 
N & 20-03-22 16:39 & 625 & 3008 & 42 & -14 & --- \\ 
N & 20-03-22 16:42 & 880 & 3011 & 42 & -15 & --- \\ 
N & 20-03-27 15:56 & B & 6010 & 40 & -6 & --- \\    
N & 20-03-27 15:46 & 550 & 3002 & 39 & -4 & --- \\  
N & 20-03-27 15:49 & 625 & 3002 & 39 & -4 & --- \\  
N & 20-03-27 15:52 & 880 & 3010 & 39 & -5 & --- \\  
N & 20-04-03 15:16 & 550 & 3010 & 39 & +3 & --- \\  
N & 20-04-03 15:19 & 625 & 3007 & 39 & +3 & --- \\  
N & 20-04-03 15:22 & 880 & 3002 & 39 & +2 & --- \\  
N & 20-04-08 16:43 & B & 6003 & 52 & -11 & --- \\   
N & 20-04-08 14:21 & 550 & 3002 & 37 & +14 & --- \\ 
N & 20-04-08 14:23 & 625 & 3004 & 37 & +14 & --- \\ 
N & 20-04-08 14:26 & 880 & 3006 & 37 & +13 & --- \\ 
N & 20-04-17 16:00 & B & 6002 & 50 & -2 & --- \\    
N & 20-04-17 15:50 & 550 & 3007 & 49 & -0 & --- \\  
N & 20-04-17 15:53 & 625 & 3002 & 49 & -1 & --- \\  
N & 20-04-17 15:56 & 880 & 3003 & 50 & -1 & --- \\  
N & 20-04-24 15:43 & 550 & 3003 & 52 & +3 & --- \\  
N & 20-04-24 15:41 & 625 & 3003 & 52 & +3 & --- \\  
N & 20-04-24 15:35 & 880 & 3004 & 51 & +4 & --- \\ \end{tabular}
\\
\end{tabular}
\end{center}
\end{small}
\end{table}

\ru{\section{Влияние конечности углового размера звезды при восстановлении изображения}}
\en{\section{The effect of finite angular size of the star on the image reconstruction}}
\label{app:finitesize}

\ru{
Представим наш объект как сумму источника неполяризованного источника малых размеров и протяженной поляризованной оболочки. Запишем Фурье--образы распределений параметров Стокса для такого объекта (здесь и далее в этом разделе зависимость от пространственной частоты опущена):}
\en{
Let us decompose some object into a sum of unpolarized point--like source and polarized extended envelope. The Fourier transforms of Stokes parameters distributions for his object will be (in this appendix the dependence on spatial frequency is omitted):}
\begin{equation}
\widetilde{I} = \widetilde{I}_\star+\widetilde{I}_\mathrm{env},\,\,\, \widetilde{Q} = \widetilde{Q}_\mathrm{env},\,\,\,\widetilde{U} = \widetilde{U}_\mathrm{env}.
\label{eq:decomp}
\end{equation}
\ru{
Малость источника неполяризованного излучения выражается в том что его видность мало отклоняется от единицы в пределах измеряемой области пространственных частот: $\widetilde{I}_\star-1 \ll 1$. Также будет разумно предположить, что вклад поляризованного излучения мал: $\widetilde{Q}_\mathrm{env}\ll\widetilde{I}_\star$ и $\widetilde{U}_\mathrm{env}\ll\widetilde{I}_\star$.}
\en{
Unpolarized source visibility does not deviate significantly from unity in the probed frequency domain: $\widetilde{I}_\star-1 \ll 1$. Also it is reasonable to assume that the fraction of polarized radiation is small: $\widetilde{Q}_\mathrm{env}\ll\widetilde{I}_\star$ and $\widetilde{U}_\mathrm{env}\ll\widetilde{I}_\star$.
}

\begin{figure}[h]
\centering
\includegraphics[width=160mm]{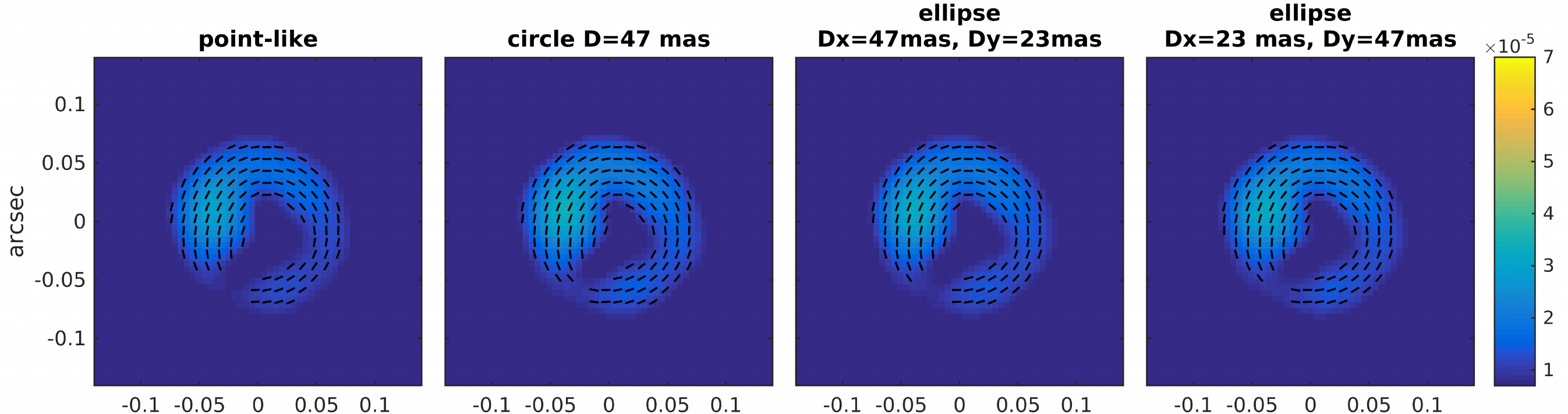}
\begin{small}
\caption{\ru{Изображение околозвездной оболочки Бетельгейзе, восстановленное методом, описанным в Приложении~\ref{app:finitesize} (дата 2019-12-11, полоса 550 нм). Слева направо, источник неполяризованного излучения: 1) точечный 2) равномерно освещенный диск диаметром 38~mas 3) равномерно освещенный эллипс размерами 38~mas по $x$ и 19~mas по $y$ 4) такой же эллипс, но повернутый на $90^{\circ}$.}
\en{
The image of circumstellar envelope of Betelgeuse in polarized light reconstructed by the method described in Appendix~\ref{app:finitesize} (date 2019-12-11, band 550 nm). From left to right, the source of unpolarized radiation is: 1) point--like 2) uniform disk with diameter of 47~mas 3) uniform ellipse with size of 47~mas along $x$ axis and 23~mas along $y$ axis 4) the same ellipse, but rotated by $90^{\circ}$.
}
\label{fig:finitesize}}
\end{small}
\end{figure}

\ru{
Cконструируем выражения следующего вида:}
\en{
Now we construct the following expressions:}
\begin{equation}
S_Q = \frac{1-\mathcal{R}_Q}{1+\mathcal{R}_Q},\,\,\,S_U = \frac{1-\mathcal{R}_U}{1+\mathcal{R}_U}.
\end{equation}
\ru{
Подстановка в них определений (\ref{eq:defR}) и (\ref{eq:decomp}) дает:}
\en{
Substituting here the definitions (\ref{eq:defR}) and (\ref{eq:decomp}) we get:}
\begin{equation}
S_Q = \frac{\widetilde{Q}_\mathrm{env}}{\widetilde{I}_\star+\widetilde{I}_\mathrm{env}},\,\,\,S_U = \frac{\widetilde{U}_\mathrm{env}}{\widetilde{U}_\star+\widetilde{U}_\mathrm{env}}
\end{equation}
\ru{
Учитывая, что знаменатель в этих выражениях мало отличается от единицы, можно оценить Фурье-образы распределений Стоксов $Q$ и $U$ и затем, обращая преобразование Фурье, восстановить изображение оболочки в поляризованной интенсивности. Пример результата такого восстановления приведен на левой панели Рис.~\ref{fig:finitesize}.}
\en{
Taking into account that the denominator in these expressions is close to unity, it is possible to estimate Fourier transforms of Stokes distribution $Q$ and $U$ and then estimate the image of the envelope in polarized light by taking inverse Fourier transforms. 
}

\ru{
Бетельгейзе по оценкам \citet{Michelson1921} имеет угловой диаметр 47~mas в видимом диапазоне, что уже сравнимо с дифракционным разрешением нашего инструмента: $\lambda/D\approx50$~mas. Результат восстановления изображения, в предположении что $\widetilde{I}_\star$ соответствует равномерно освещенному диску диаметром 38~mas, приведен на Рис.~\ref{fig:finitesize}.
}
\en{
According to \citet{Michelson1921} the angular diameter of Betelgeuse is 47~mas in visible, which is already comparable with diffraction limited resolution of our instrument: $\lambda/D\approx50$~mas. The image reconstructed under assumption that $\widetilde{I}_\star$ corresponds to a uniform disk with a diameter of 47~mas is presented in Fig.~\ref{fig:finitesize}.
}

\ru{
На этом же рисунке даны изображения восстановленные для тех случаев когда звезда представляет собой эллипс, сплюснутый в два раза по оси $x$ и по оси $y$. Эта ситуация призвана смоделировать вид фотосферы из работы \citet{Montarges2020}. Как видно, во всех случаях учет конечности углового размера неполяризованного источника не оказывает значимого влияния на результирующее изображение. Поэтому в данной работе мы предполагаем что $\widetilde{I}_\star=1$.
}

\en{
The same figure contains images reconstructed for the case of elliptical stellar image compressed 2 times along $x$ and $y$ axes. This situation models the appearance of photosphere from \citep{Montarges2020}. As one can see, in all the cases taking into account the finite angular size of the star does not affect much the resulting image. Therefore in this work we assume that $\widetilde{I}_\star=1$.
}

\end{document}